\documentclass{svproc}
\usepackage{amsmath}
\usepackage{graphicx}
\usepackage{url}

\begin{document}
\mainmatter             
\title{Microscopic optical potential from chiral effective field theory}

\author{T. R. Whitehead*, Y. Lim, and J. W. Holt}

\authorrunning{T. R. Whitehead, Y. Lim, and J. W. Holt} 

\institute{Cyclotron Institute, Texas A\&M University, College Station, TX 77843, USA 
\and
Department of Physics and Astronomy, Texas A\&M University, College Station, TX 77843, USA
\newline
$^*$twhitehead@tamu.edu}

\maketitle

\begin{abstract}
We formulate a microscopic optical potential from chiral two- and three-body forces. The real and imaginary central terms of the optical potential are obtained from the nucleon self-energy in infinite matter, while the real spin-orbit term is extracted from a nuclear energy density functional constructed from the density matrix expansion using the same chiral potential. The density-dependent optical potential is then folded with the nuclear density distributions for selected Calcium isotopes resulting in energy-dependent nucleon-nucleus optical potentials from which we study proton-nucleus elastic scattering cross sections calculated using the TALYS reaction code. We compare the results of the microscopic calculations to phenomenological models and experimental data.
\end{abstract}

\section{Chiral nuclear optical model potentials}

Optical model potentials are widely used to predict nucleon-nucleus scattering cross sections and reaction observables by replacing the complicated many-body system of nucleons interacting through two- and three-body forces with an average complex and energy-dependent single-particle potential. Phenomenological models \cite{kd03} fitted to experimental data are very successful at describing scattering processes for nuclei near stability, but high-quality microscopic optical potentials may be more reliable for reactions involving exotic isotopes for which experimental data are scarce. Recently, microscopic optical potentials in homogeneous nuclear matter have been constructed \cite{chiralop,chiralop2} based on realistic chiral two- and three-body forces. The aim of the present work is to extend this description to the case of finite nuclei, with a special focus on proton elastic scattering off calcium isotopes.

In quantum many-body theory, the nuclear optical potential is identified with the nucleon self-energy.  We begin by computing the nucleon self-energy in infinite homogeneous nuclear matter at a given density and isospin asymmetry starting from a realistic chiral nuclear interaction \cite{coraggio14} with momentum-space cutoff $\Lambda = 450$\,MeV. The real and imaginary central terms of the optical potential arise naturally when the nucleon self-energy is computed to second order in many-body perturbation theory. The real spin-orbit term cannot be extracted from nuclear matter calculations and in the present work is instead calculated from the Negele-Vautherin density matrix expansion \cite{edf} using the same chiral potential. The density-dependent optical potential is then folded with the relevant nuclear density distribution for the isotope under investigation, calculated using a Skyrme effective interaction fitted to the nuclear equation of state derived from the same chiral potential. The result is an energy-dependent nucleon-nucleus optical potential in position space. This local density approximation (LDA) \cite{jlm} is known to give a poor description of the optical potential surface diffuseness, and therefore in the present work we employ an improved local density approximation (ILDA) that accounts for the non-zero range of the nuclear force:
\begin{equation}
\label{eq:ilda}
U(E,r)_{ILDA}=\frac{1}{(t\sqrt{\pi})^3}\int U(E,r') e^{\frac{-|\vec{r}-\vec{r}'|^2}{t^2}} d^3r',
\end{equation}
where $t$ is a distance scale associated with the average range of the nucleon-nucleon interaction. In the present study we vary $t$ within the range $1.15\,{\rm fm} < t < 1.25$\,fm.

\section{Results}
\begin{figure}[t]
\begin{center}
\includegraphics[scale=0.27]{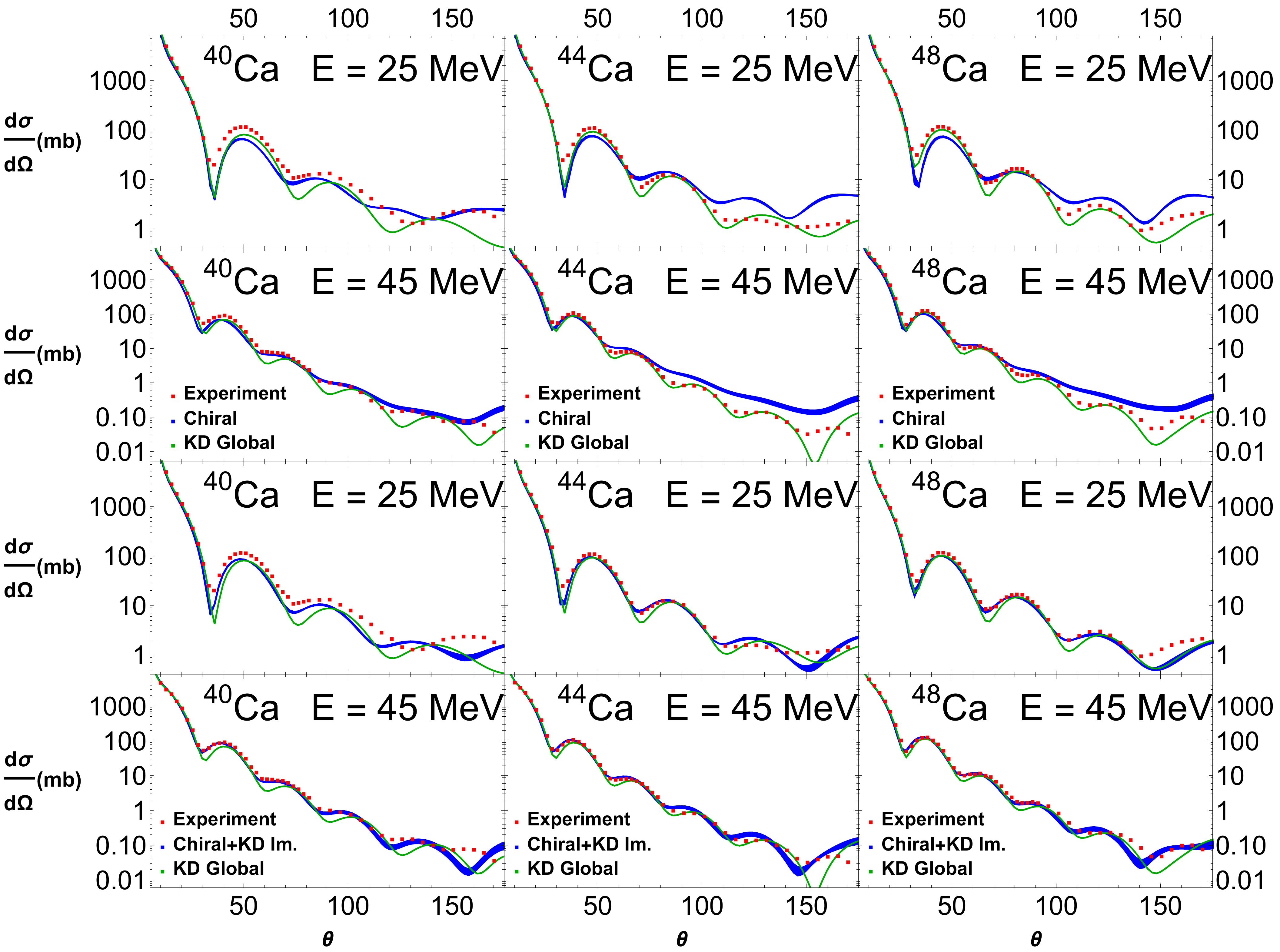}
\caption{Differential elastic scattering cross sections for proton-nucleus scattering at incident energies $E=25,45$\,MeV. We show results from microscopic chiral nuclear forces (blue), the phenomenological Koning-Delaroche (KD) optical potential (green), and experimental data (red). In the bottom two rows, we have replaced the microscopic imaginary part with the phenomenological imaginary part of the KD optical potential. }
\label{csplot1}
\end{center}
\end{figure}

We have implemented the nuclear optical potentials described above in the TALYS reaction code \cite{talys}.
In the top two rows of Fig.\ \ref{csplot1} we plot the proton-nucleus differential elastic scattering cross sections at the two energies $E=25,45$\,MeV for the isotopes $^{40}$Ca, $^{44}$Ca, and $^{48}$Ca. Experimental data are shown as the red points, while the results from the microscopic optical potentials are shown with the blue band. The uncertainties giving rise to the theoretical error band are obtained by varying the ILDA range parameters for the both the central and spin-orbit components. In the future we plan to estimate also the uncertainties arising from the choice of chiral potential by varying the momentum-space cutoff, the order in the chiral expansion, and the regulating function. We also show in the top two rows of Fig.\ \ref{csplot1} the results (green curves) from the Koning-Delaroche phenomenological optical potential as it is implemented in the TALYS reaction code. We see that the microscopic optical potentials from chiral effective field theory give an overall reasonable description of the elastic scattering cross section within the chosen energy regime. However, at higher energies and larger scattering angles, the description starts to deteriorate.

Comparing the real and imaginary components of the microscopic optical potential to those from phenomenology, we find excellent agreement in the real part but the microscopic imaginary optical potential is too strongly absorptive. This feature is ubiquitous in nuclear matter optical potential calculations and is the reason why semi-microscopic optical potentials used today implement energy-dependent strength factors \cite{goriely07}. To test this, we show in the bottom two rows of  Fig. \ref{csplot1}, the results from our microscopic optical potentials for which the imaginary part has been replaced by that from the KD optical potential. We see that this replacement dramatically improves the differential elastic scattering cross sections across all energies and target isotopes. In the future we plan to investigate higher-order perturbative contributions to the self-energy and their effect on the imaginary part of the optical potential.

Work supported by the U.S. Department of Energy National Nuclear Security Administration under Grant No. DE-NA0003841 and by the National Science Foundation under Grant No.\ PHY1652199.

\end{document}